%% file: RASANEN_Sami_ICFNP2015.tex
\documentclass[epj]{webofc}
\usepackage[varg]{txfonts}   
\wocname{EPJ Web of Conferences}
\woctitle{ICNFP 2015}
\include{defs}

\begin{document}
\selectlanguage{english}
\title{ALICE overview}
%
%

\author{Sami~S.~R\"as\"anen\inst{1,2}\fnsep\thanks{\email{sami.s.rasanen@jyu.fi}} on behalf of the ALICE Collaboration
}

\institute{University of Jyvaskyla, Department of physics, P.O. Box 35, FI-40014, University of Jyvaskyla, Finland
\and
           Helsinki Institute of Physics, P.O.Box 64, FI-00014 University of Helsinki, Finland
}

\abstract{%
  Recent results from the ALICE experiment are presented with a particular emphasis on particle identification, the nuclear modification factor ($\raa$) and azimuthal anisotropy ($v_2$). Comparison of lead-lead and proton-lead results reveals evidence of collectivity in small systems.
}
\maketitle
%
\section{Introduction}\label{intro}

The main goal of ultrarelativistic heavy ion collisions is to study the thermal properties of quantum chromodynamics (QCD). It is widely accepted that the thermal state of QCD, the quark-gluon plasma (QGP), is reached in high energy nucleus-nucleus collisions at BNL-RHIC and CERN-LHC. Properties of the QGP medium, such as shear viscosity to entropy ratio, existence and location of the QCD critical point, electromagnetic radiation and energy loss provide a focus for active measurements and theory development; see e.g. the Hot QCD White Paper \cite{Akiba:2015jwa}. Recently there has been a growing interest in possible collective phenomena in smaller collision systems \cite{smallSystems}, like deuteron-gold \cite{Adare:2014keg,Adamczyk:2015xjc}, helium-gold \cite{Adare:2015ctn}, proton-lead \cite{Abelev:2012ola,Aad:2012gla,Chatrchyan:2013nka} or even in proton-proton \cite{Khachatryan:2010gv,Aad:2015gqa,Khachatryan:2015lva}.

The ALICE experiment \cite{Aamodt:2008zz} is the dedicated heavy ion experiment at the LHC. Compared to ATLAS \cite{Aad:2008zzm} and CMS \cite{Chatrchyan:2008aa}, ALICE is a low luminosity experiment with more limited acceptance but equipped with excellent particle identification (PID) and tracking capabilities down to very low transverse momentum, \pt{}.
\begin{figure}[h]
\centering
\sidecaption
\includegraphics[width=7.0cm,clip]{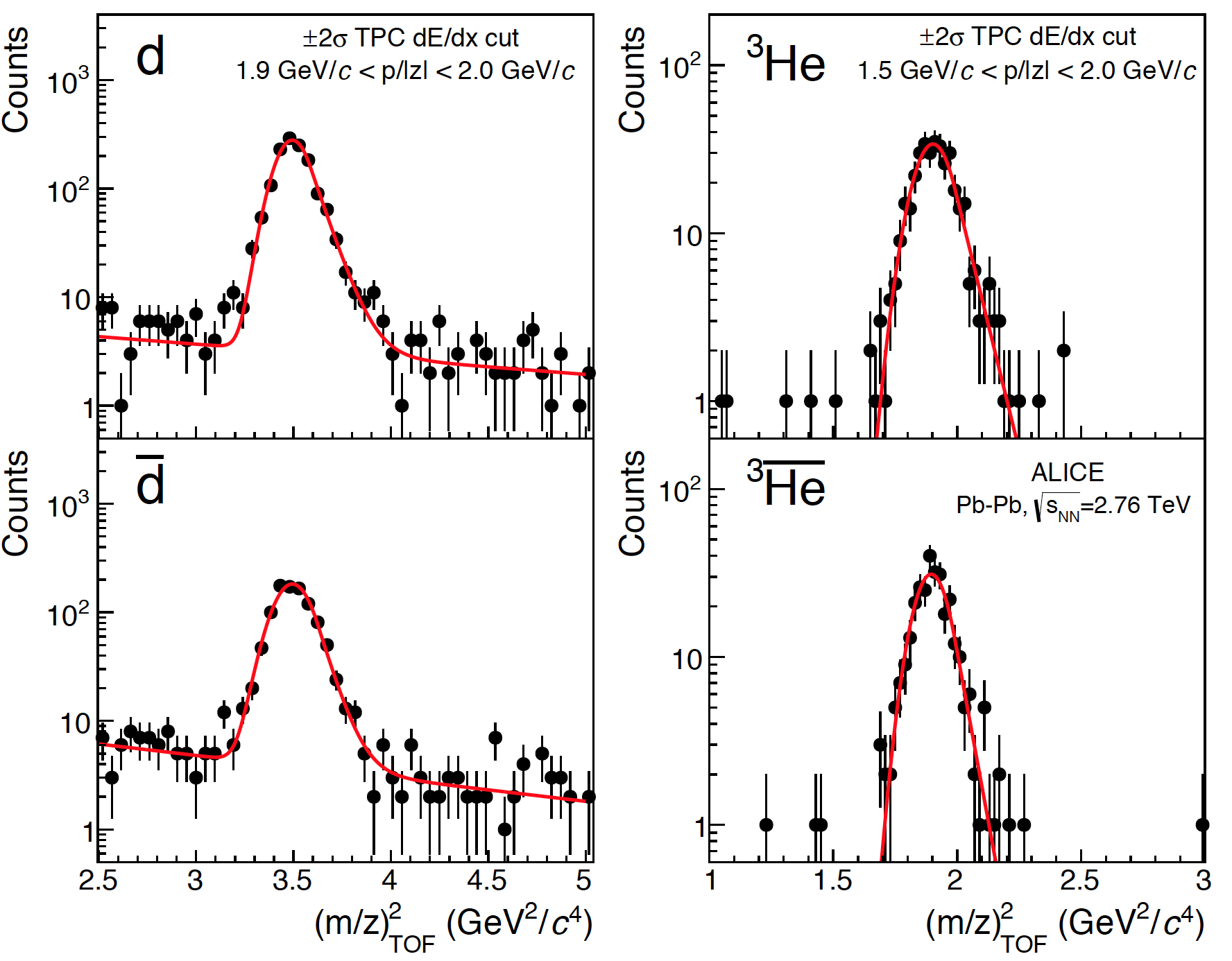}
\includegraphics[width=7.0cm,clip]{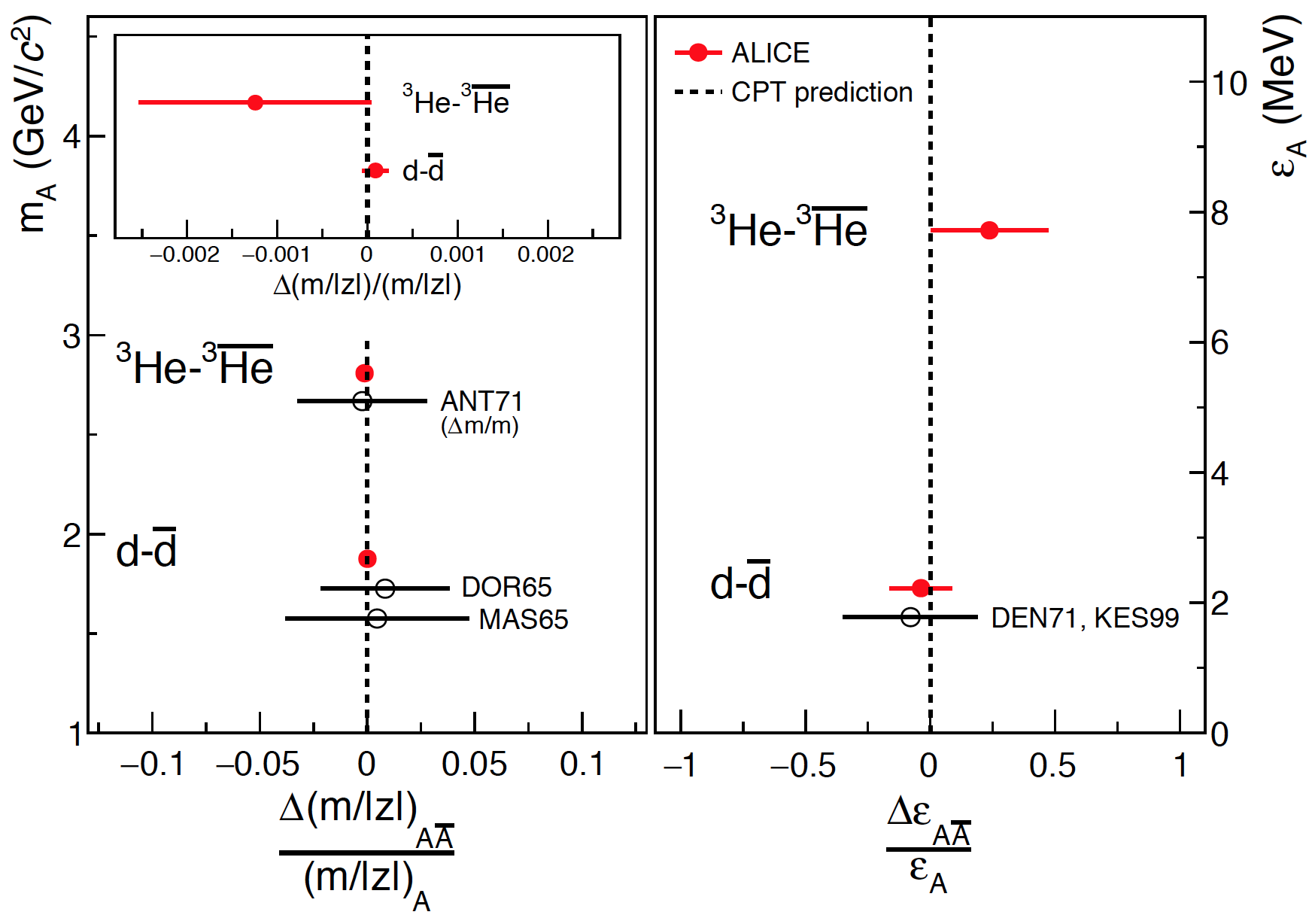}
\caption{Left: The ALICE measurements \cite{Adam:2015pna} for squared mass-over-charge ratio distributions for deuterons (left) and $^{3}$He (right) in selected rigidity intervals. Particle and anti-particle spectra are in the top and bottom plots, respectively. Right: Results for d-d and $^{3}$He-$^{3}$He mass-over-charge ratio and binding energy differences compared with CPT invariance expectation (dotted lines) and other measurements. }
\label{fig:lightIons}
\end{figure}
One example of the performance of the ALICE tracking is a recent measurement of the mass difference between light nuclei and anti-nuclei \cite{Adam:2015pna}. The left panel of \fig{fig:lightIons} shows observed counts with respect to the squared mass-over-charge ratio in the \pbpb\ collisions at $\sqrts=2.76\gev$ for selected rigidity ($p/|z|$) intervals and $2\sigma$ cut of for TPC $dE/dx$ expectations for (anti-)deuterons and (anti-)helium-3. The mass of the light nuclei can be measured very precisely by fits to these distributions. The right panel of \fig{fig:lightIons} shows the resulting mass differences and binding energy differences compared to earlier measurements. If the Charge conjugation-Parity-Time reversal (CPT) --symmetry is exact, then these differences should be zero. ALICE has now presented the most precise measurement confirming CPT invariance to hold in relative mass differences at 0.1\% level, which gives stringent constraints to any effective theory where CPT is broken.

On top of the measurements of heavy ion collisions, ALICE has also a proton-proton program based upon the strengths of the experiment. One general motivation is to test and tune event generators and basic perturbative QCD predictions, which is important in its own right,but also a precise understanding of QCD background is often crucial for the electroweak measurements at the LHC. Some recent measurements that demonstrate the PID capabilities of ALICE include identified hadrons and particle ratios at $\sqrts=7\tev$ \cite{Adam:2015qaa} and $\pi^0$'s at $\sqrts=2.76\tev$ \cite{Abelev:2014ypa}. ALICE has also measured (charged) jet spectrum at $\sqrts=7\tev$ \cite{ALICE:2014dla} and found a good agreement with ATLAS. Here, the excellent tracking capability of ALICE was used to measure the mean number the mean number of constituents down to low-\pt{}\ jets. On top of these examples, the obvious and important part of the ALICE \pp\ program is to measure reference spectra to the heavy ion collisions.

The rest of this paper will concentrate on summarizing recent ALICE results on the nuclear modification factor $\raa$ and elliptic flow $v_2$ with an emphasis on particle identification and small systems.

\section{Nuclear modification factor}\label{sec:RAA}

It has long been predicted that partons traversing hot QCD matter would lose energy and attenuate, but such a phenomena was not observed in early experiments at the CERN SPS \cite{Wang:1998hs}. A suppression of high transverse momentum hadron was first observed by experiments at RHIC \cite{Adcox:2001jp,Adam:2015qaa}. To quantify the nuclear modification, PHENIX experiment introduced the nuclear modification factor \cite{Adcox:2001jp}
\bge
\raa(\pt{})=\frac{({\rm Yield \ in \ Pb+Pb})}{({\rm Number \ of \ collisions})\times({\rm Yield \ in \ p+p})}
    = \frac{dN^{AA}/d\pt{}d\eta}{\mean{N_{\rm bin}}dN^{\rm{pp}}/d\pt{}d\eta}.
\ende
If the nucleus-nucleus collision could be described as a superposition of independent proton-proton collisions, then \raa\ would equal unity for all \pt{}. At the low-\pt{}\ $\raa<1$ since the soft particle production in heavy ion collisions scales with the number of participants rather than the number of collisions \cite{Aamodt:2010cz}. At high-\pt{}\ at the SPS $\raa>1$ which is connected with Cronin enhancement \cite{Antreasyan:1978cw}. At RHIC a suppression with respect to binary scaling was observed at the high-\pt{}.

In proton-lead collisions, the corresponding nuclear modification factor \rpa\ measures the magnitude of cold nuclear matter effects such as a modification of the nuclear parton distributions \cite{Eskola:2009uj}. As will be seen, typically the cold nuclear effects are small at very high energy. To quantify nuclear modifications precisely, the most favourable case is when all collision systems, \pp, \ppb\ and \pbpb, are measured at the same collision energy and by the same detector. Currently at the LHC, there is a measured \pp\ reference for the $\sqrts=2.76\tev$ lead-lead measurements, but not yet for the $\sqrts=5.02\tev$ proton-lead run. This optimal scenario will be achieved at the end of 2015 when the LHC will provide both $\sqrts=5.02\tev$ reference \pp\ run and also \pbpb\ collisions at this same energy.
 
Following subsections discuss the \raa\ of various single particle spectra. ALICE has also measured \raa\ of jets \cite{Adam:2015ewa}. See also the dedicated talk on jets in heavy ions collisions presented at this conference \cite{jetTalk} .

\subsection{Light flavours}\label{sec:RAA-LF}

The first ALICE \raa\ measurement for charged hadrons was published in \cite{Aamodt:2010jd} and immediately made an important extension to RHIC results (not shown here). RHIC results reached up to $\pt{}\sim10\gevc$ and it was open weather the data flattens in that region or if \raa\ starts to rise again. With the LHC data extending to $50\gevc$ (ALICE) or 100+\gevc\ (ATLAS \cite{Aad:2015wga}, CMS \cite{CMS:2012aa}), it is clear that \raa\ is again rising towards unity. This rules out a very simple phenomenological description of energy loss where every parton would loose a constant fraction of its energy, and also gives more constraints to detailed calculations.

\begin{figure}[h]
\centering
\sidecaption
\includegraphics[width=12.0cm,clip]{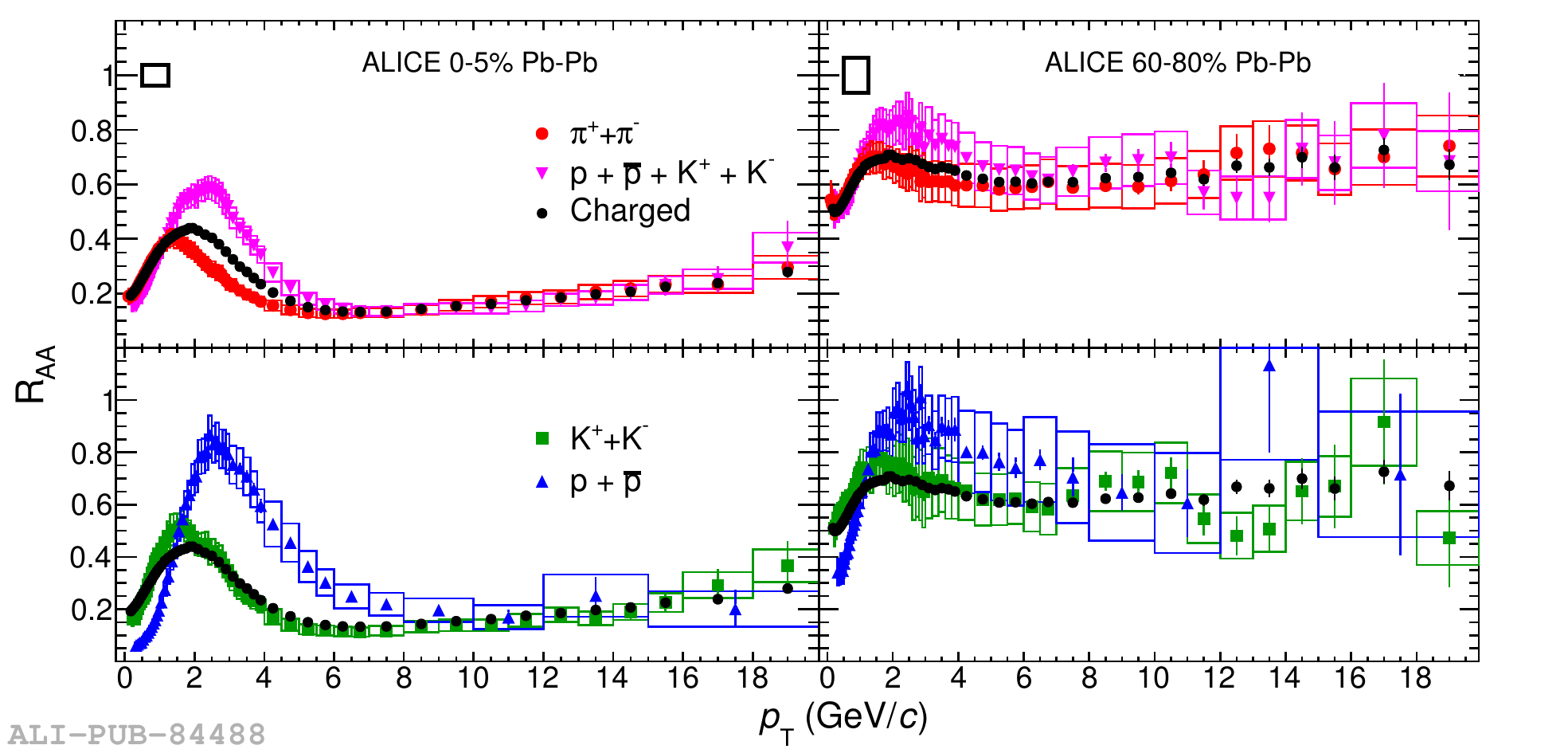}
\caption{The nuclear modification factor \raa\ as a function of \pt{} for different particle species for central (left) and peripheral (right) collisions \cite{Adam:2015kca}.}
\label{fig:RAA-LF}
\end{figure}
\fig{fig:RAA-LF} shows recent ALICE measurements of \raa\ for identified light hadron flavours; charged pions and kaons and (anti-)protons together with all charged hadrons up to $\pt{}=20\gevc$ in central (left) and peripheral (right) collisions. In central collisions, \raa\ is observed to be the same for all particle species for $\pt{}\,\gsim\,10\gevc$ suggesting that there is no direct interplay between the energy loss in the medium and the particle species composition in the hard core of the quenched jet and may disfavour models that predict significant particle species dependence also at high-\pt{}. The region $\pt{}\,\lsim\,2\dots2.5\gevc$ is generally associated with the thermal production which roughly scales as a number of participants. In the mid-\pt{}\ region, $3<\pt{}<10\gevc$, protons are clearly less suppressed than pions or kaons. Further studies are needed in order to determine whether models containing only hydrodynamics and jet quenching can describe also the intermediate \pt{}\ region or whether a different hadronization model, such as recombination \cite{Fries:2003kq}, act in this region.

\subsection{Heavy flavours}\label{sec:RAA-HF}

Dedicated heavy flavour (HF) theory \cite{HFtheory} and experimental \cite{HFexp} talks were delivered in this conference. Here, we summarize briefly ALICE results for HF~\raa. One of the key motivations to study HF in nucleus-nucleus collisions have been the expectation of reduced energy loss in the medium due to so called dead cone effect \cite{Dokshitzer:2001zm}, that arises from suppression of the forward QCD scattering amplitude due to heavy quark mass in the propagator.

\begin{figure}[h]
\centering
\sidecaption
\includegraphics[width=8.0cm,clip]{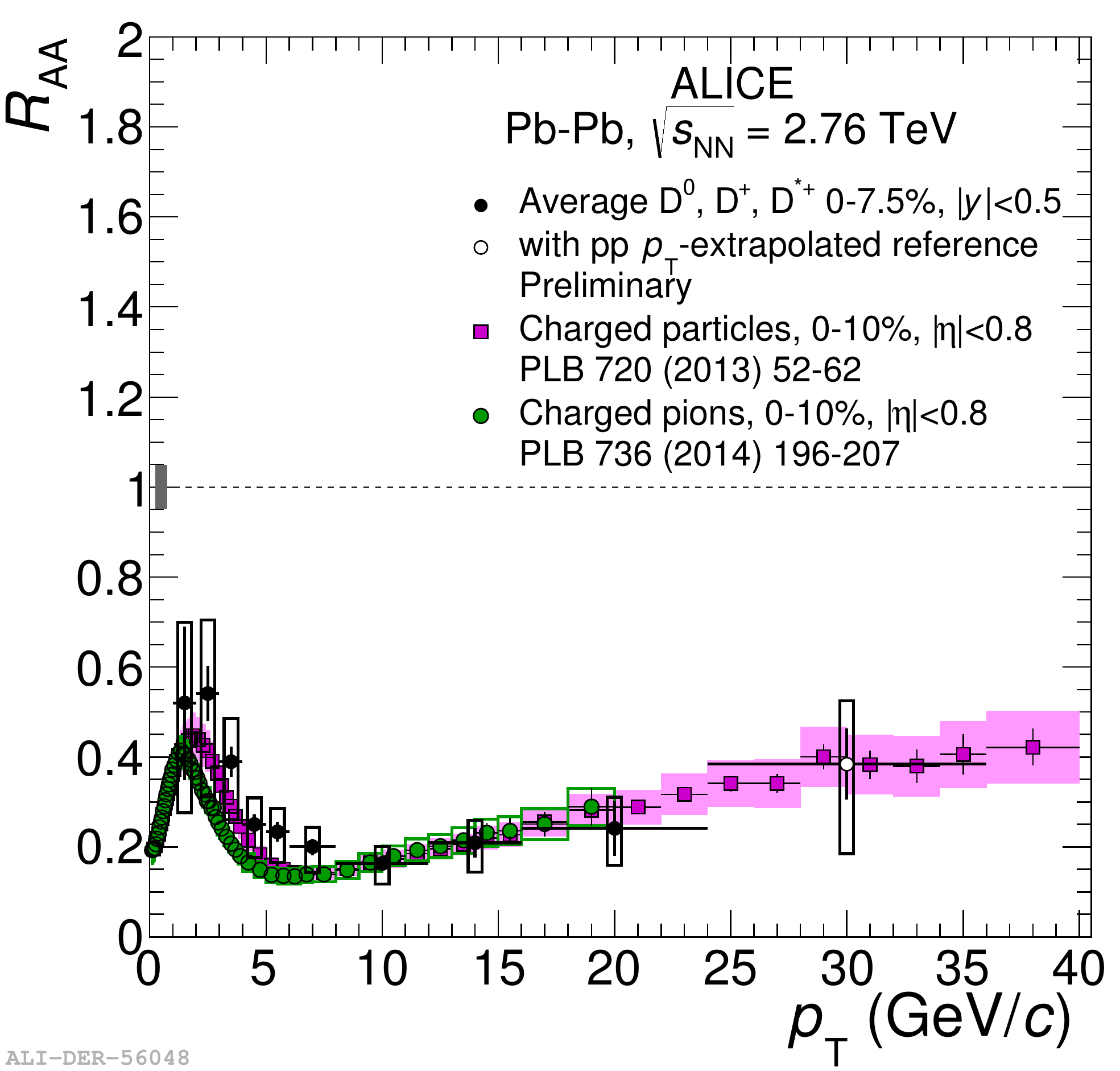}
\caption{Average D-meson $\raa$ compared to charged hadrons and - pions.}
\label{fig:RAA-D}
\end{figure}

\fig{fig:RAA-D} shows the average D-meson \raa\ compared to all charged particles and charged pions. Particularly at high-\pt{}\ the results agree within one another, and also in lower \pt{}\ within the experimental uncertainties, so the dead cone effect is not obvious. Although some models can reproduce this, they tend to have difficulties in simultaneously reproducing the elliptic flow, as will be discussed later in \sec{sec:v2-PbPb}.

Suppression of the \jpsi\ states in presence of QGP was predicted in 1986 \cite{Matsui:1986dk} and it has been studied extensively at the SPS \cite{Arnaldi:2009ph} and RHIC \cite{Adare:2007gn}. On top of the effects coming from the hot deconfined medium, there may be cold nuclear matter effects like modification of parton distribution functions or final state absorption in the nuclear medium. Both SPS and RHIC experiments found  suppression beyond any cold matter effects. At the LHC, CMS has observed a sequential suppression of the higher mass bottomonium states \cite{Chatrchyan:2012lxa} showing clear medium effects in the suppression.

An earlier measurement in ALICE \cite{Abelev:2012rv} revealed an interesting surprise: the \pt{}-integrated \raa\ of the \jpsi\ was less suppressed in central lead-lead collisions at the LHC as compared to central gold-gold collisions at RHIC. This is counter intuitive in a sense that naively one would expect the larger and more hot medium to give larger suppression.

\begin{figure}[h]
\centering
\sidecaption
\includegraphics[width=7.0cm,clip]{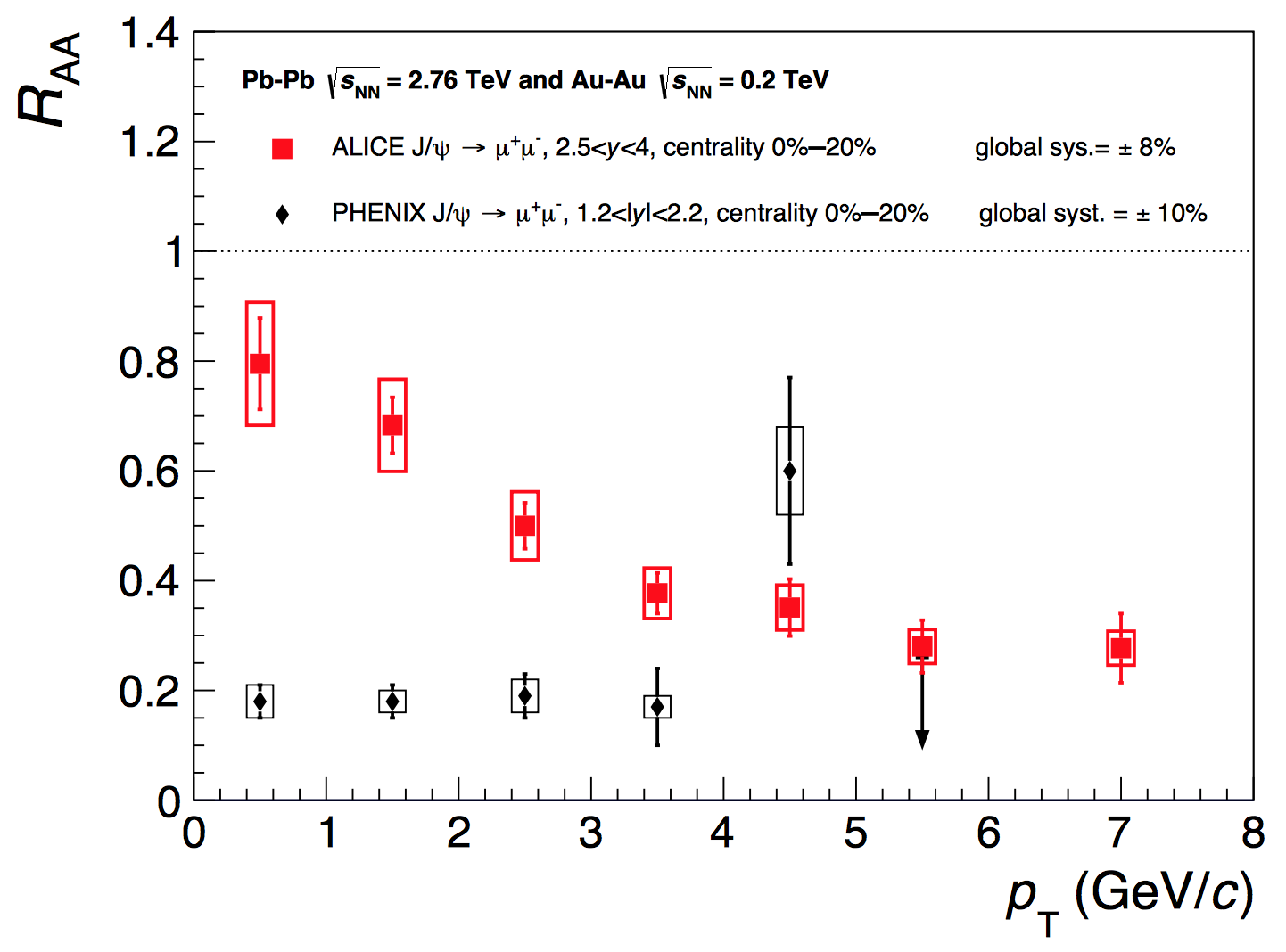}
\includegraphics[width=7.0cm,clip]{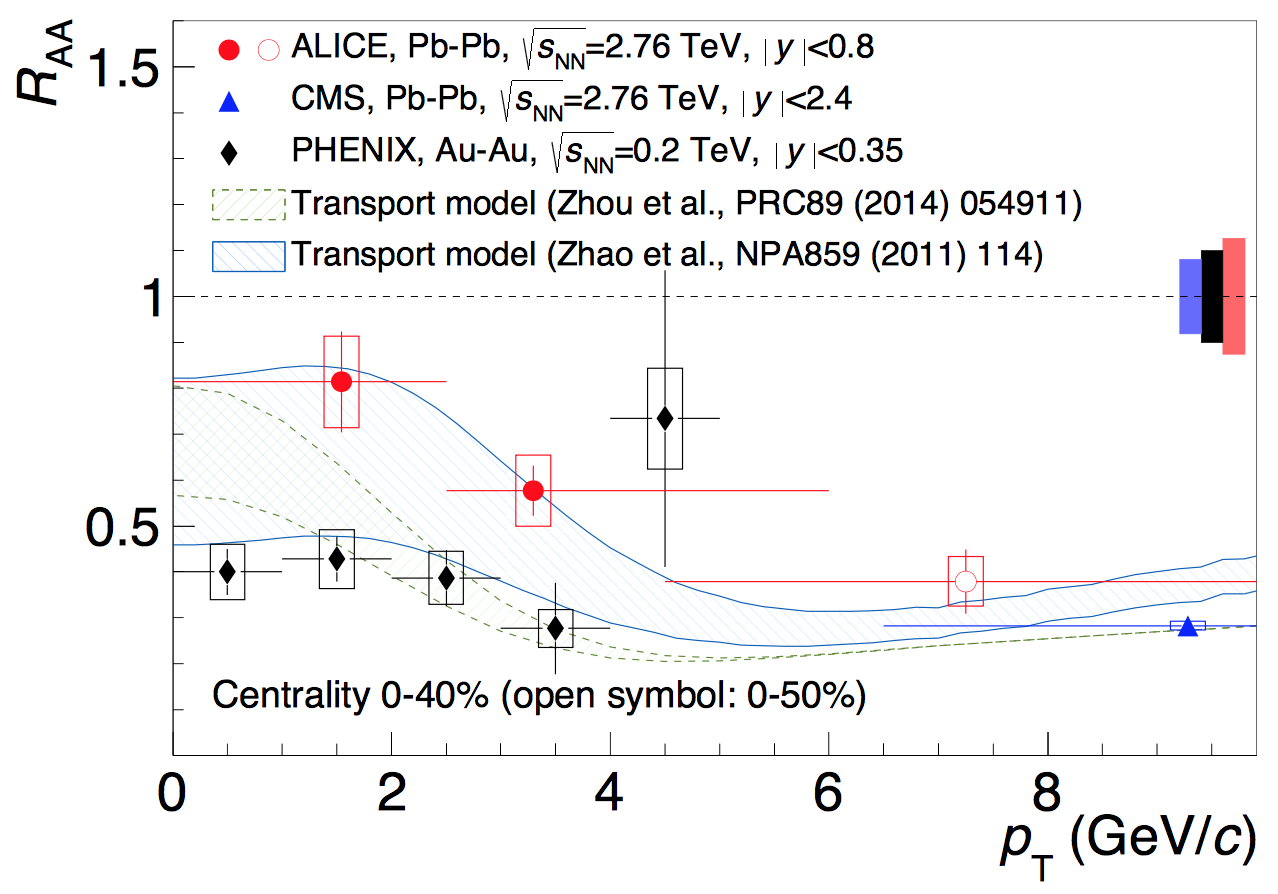}
\caption{Nuclear modification factor \raa\ of \jpsi\ in 0--20~\% most central lead-lead collisions at forward rapidity \cite{Abelev:2013ila} (left) and in 0--40~\% most central collisions at mid-rapidity \cite{Adam:2015rba} (right).}
\label{fig:RAA-JPsi}
\end{figure}
\fig{fig:RAA-JPsi} shows ALICE results for \pt{}-dependence of \jpsi\ nuclear modification factor \raa\ in central lead-lead collisions measured at forward rapidity (left panel) using di-muons pairs \cite{Abelev:2013ila} and in mid-rapidity (right panel) using di-electron pairs \cite{Adam:2015rba}. While in \cite{Abelev:2012rv,Abelev:2013ila} one compared forward LHC to mid-rapidity RHIC measurements, later measurement \cite{Adam:2015rba} confirmed that the observed smaller suppression at the LHC does not come from a trivial rapidity dependence.  Note that the last bin in the right panel has open marker indicating that the statistics was too low to determine the mean \pt{}\ in that bin, hence the marker is placed at the bin center.

A possible explanation for smaller suppression comes from higher density of charmed quarks at the LHC that would increase the recombination probability to \jpsi. The right panel in \fig{fig:RAA-JPsi} shows results from two transport models \cite{Zhou:2014kka,Zhao:2011cv} that can describe the rise of \raa\ towards the small \pt{}. In both models the rise comes from recombination.

\subsection{Proton-lead collisions}\label{sec:RAA-pPb}

In the early RHIC era, deuteron-gold collisions were considered mainly as a control for heavy ions. It was expected that there is no medium created and the measurements would probe only cold nuclear matter effects. Originally proton-lead collisions were expected to play the same role also at the LHC. Experiments at both accelerator centers have now found that there may be collective behaviour also in small systems, as will be discussed in \sec{sec:flow-pPb}.

\begin{figure}[h]
\centering
\sidecaption
\includegraphics[width=7.0cm,clip]{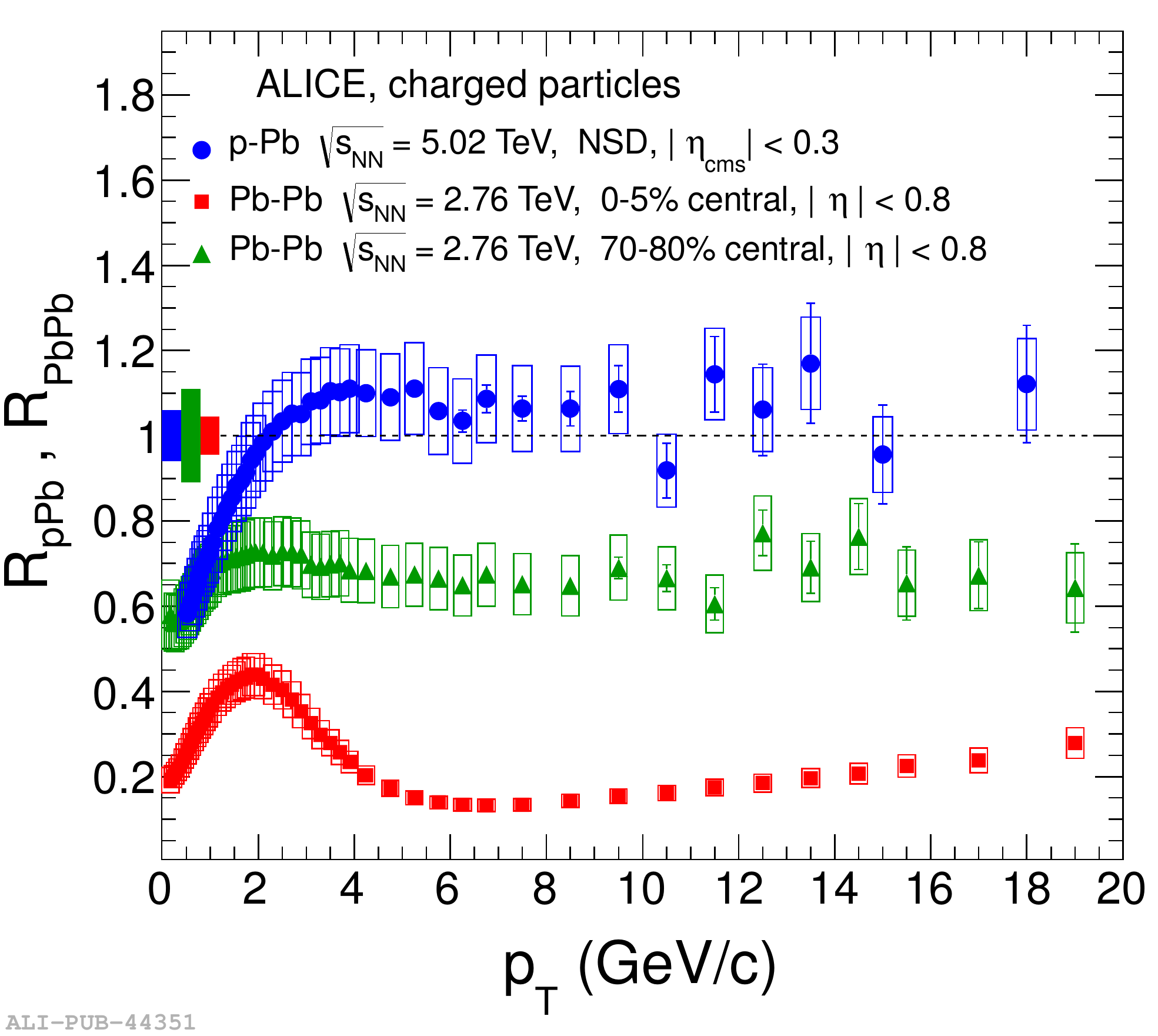}
\includegraphics[width=7.0cm,clip]{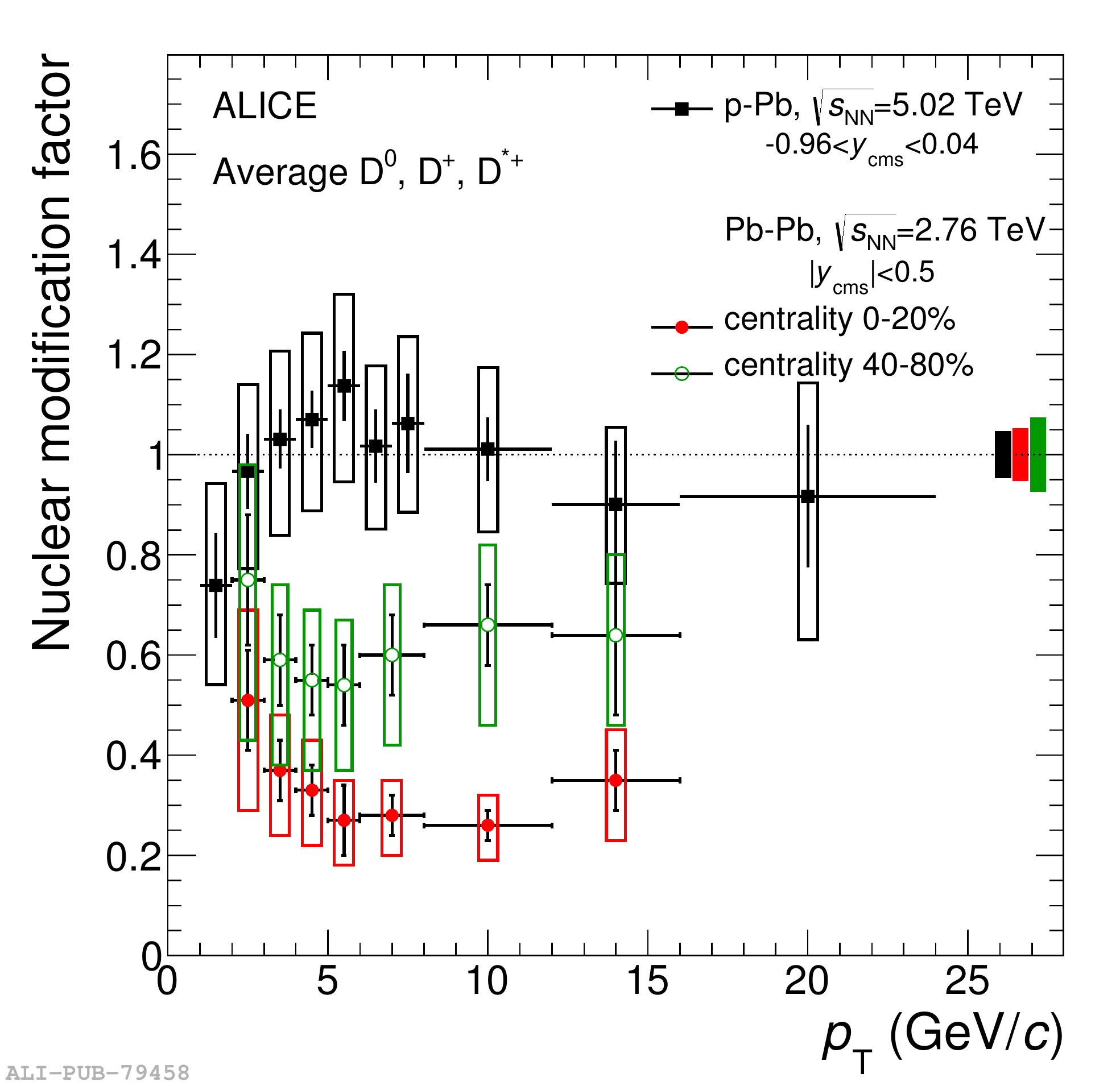}
\caption{Nuclear modification factor \raa\ for charged hadrons \cite{ALICE:2012mj} (left) and average D-mesons \cite{Abelev:2014hha} (right) in minimum bias proton-lead collisions compared to central and peripheral lead-lead collisions.}
\label{fig:RAA-pPb}
\end{figure}
\fig{fig:RAA-pPb} shows the nuclear modification factor \raa\ for charged hadrons \cite{ALICE:2012mj} (left) and average D-mesons \cite{Abelev:2014hha} (right) in minimum bias proton-lead collisions compared to central and peripheral lead-lead collisions. At high-\pt{}, \raa\ is consistent with unity for both charged hadrons and average D-mesons in \ppb\ collisions. On the other hand, a clear suppression is seen in central heavy ion collisions. Peripheral heavy ion results lie in between indicating smooth and gradual disappearance of the nuclear effects. The observed cold nuclear matter effects (at high-\pt{}) are rather modest. This was expected if the only source of cold nuclear modifications is due to the nuclear parton distribution functions, see e.g. \cite{Helenius:2012wd}. However, saturation physics phenomenology \cite{Tribedy:2011aa} gives results that are also consistent with the data.

\section{Elliptic flow}\label{sec:v2}

If the fireball created in the relativistic heavy ion collisions thermalizes, then there is a strong hydrodynamical pressure gradient that drives the system to collective motion towards the vacuum outside the droplet. In a very simple blast wave model, the fireball is assumed to break up instantaneously, at some constant proper time, parametrized collective transverse flow profile, in which case the invariant yield of thermal particles becomes \cite{Schnedermann:1993ws}
\bge\label{eq:BW}
E\frac{dN}{d^3p}\propto m_{\rm T}\int_0^{R_A}dr \ rI_0\left(\frac{\beta_T\gamma_T\pt{}}{T_{\rm kin}}\right)K_1\left(\frac{\gamma_Tm_{\rm T}}{T_{\rm kin}}\right), \quad{\rm where} \ \beta_T=\beta_T(r)=\beta_{\rm max}\left(\frac{r}{R_A}\right)^\alpha.
\ende
The gamma-factor is related to the collective transverse flow $1/\gamma_T=\sqrt{1-\beta_T^2}$, transverse mass $m_T=\sqrt{\ptkv{}+m^2}$ and $R_A$ is the nuclear radius. The model has four parameters to be determined by a fit to the data: the kinetic freezeout temperature of the system, $T_{\rm kin}$, the maximum flow velocity, $\beta_{\rm max}$, the shape parameter for the flow profile, $\alpha$, and an overall normalization related to the size (also the lifetime) of the fireball. In the case of constant radial flow $\beta_T=\beta_0=(\rm const.)$, i.e. shape parameter $\alpha=0$, one finds that asymptotically at $\pt{}\gg \max(m,T_{\rm kin})$
\bge\label{eq:mTscaling}
\frac{1}{m_T}\frac{dN}{dm_Tdy}\sim\exp\left(\frac{-m_T}{T_{\rm eff}}\right), \quad{\rm where} \ T_{\rm eff}\equiv T_{\rm kin}\sqrt{\frac{1+\beta_0}{1-\beta_0}}.
\ende
This illustrates the blue shifts in spectra coming from the collective radial expansion. The increase of the apparent temperature is the dominant flow effect in heavy ion collisions.

\begin{figure}[h]
\centering
\sidecaption
\includegraphics[width=8.0cm,clip]{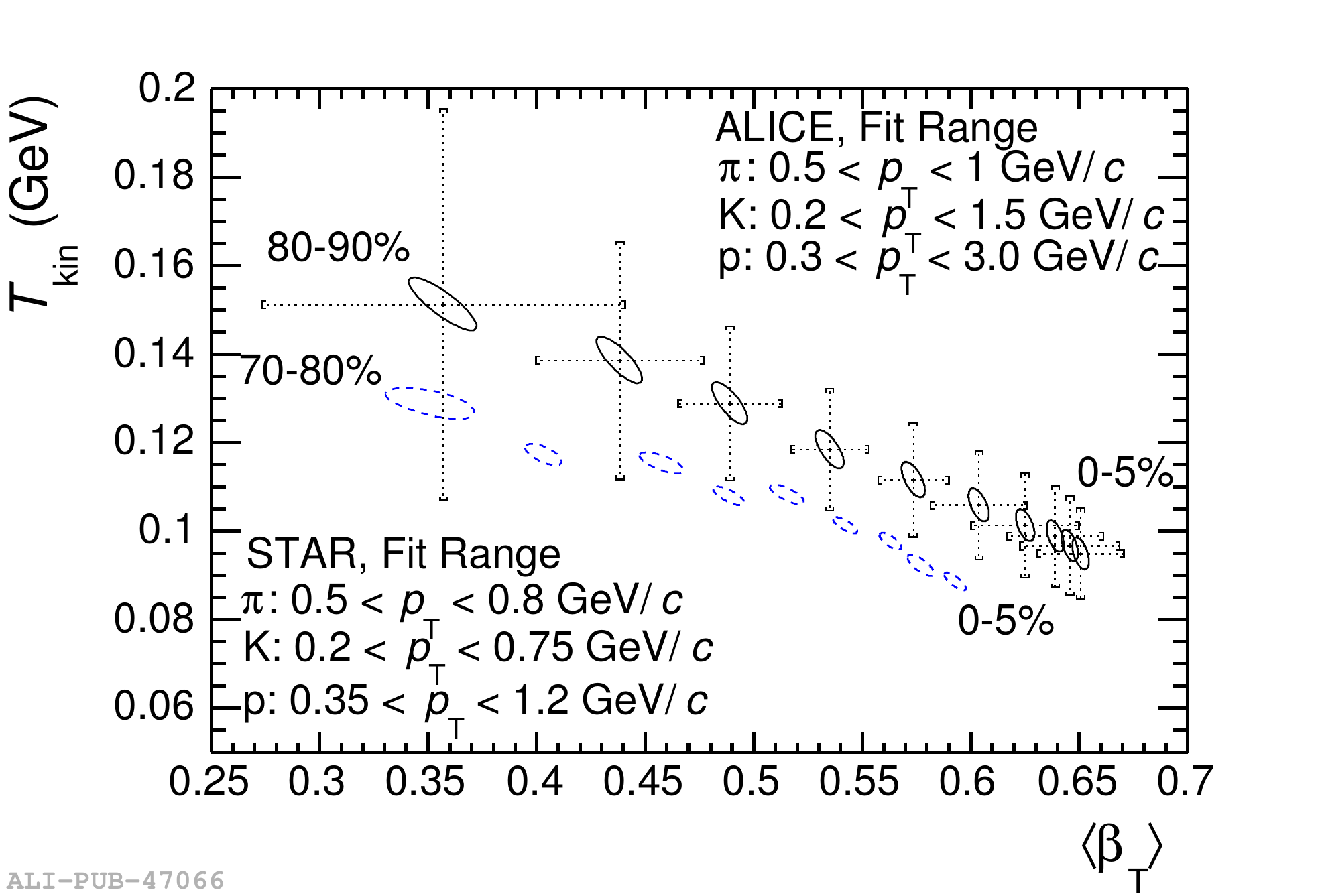}
\caption{Blast wave fit parameters break up temperature $T_{\rm kin}$ and mean transverse velocity $\mean{\beta_T}$ at the LHC and in RHIC \cite{Abelev:2013vea}. With increasing centrality, the mean velocity grows and break up temperature reduces. This can be interpreted such that in more central collisions the average lifetime of the system is longer leading to stronger flow and later break up.}
\label{fig:BW}
\end{figure}
\fig{fig:BW} shows the blast wave parameters $T_{\rm kin}$ and mean of the transverse velocity $\mean{\beta_T}$ at the LHC (solid black) and in RHIC (dashed blue) \cite{Abelev:2013vea}. Centrality increases when moving from left to right. Average flow velocity increases and freezeout temperature decreases with growing centrality. This suggests that more central collisions have longer lifetime and hence the flow has time to build up.

The blast wave model has several short comings. Perhaps the most important, it does not contain any dynamics and it neglects the fact that significant fraction of pions and protons originate from the decays of heavier resonances \cite{Eskola:2002wx}. The latter could be easily taken into account by considering the resonance decay chains in the blast wave spectra but it would still only give a qualitative description. 

Hydrodynamical modeling gives a real dynamical solution of the conservation equations and is the only known thermodynamically consistent way to treat the Equation of State of strongly interacting matter in the evolution equations. The solution of the hydrodynamical evolution equations give dynamically the local temperature, densities and collective velocity in every space-time point inside the fireball. Hence all quantitative thermodynamical characteristics should follow from detailed hydrodynamical simulations. For example, see recent results found using sophisticated theoretical frameworks like VISHNU \cite{Song:2013qma}, MUSIC \cite{Gale:2012rq} or pQCD+saturation+viscous hydrodynamics \cite{Niemi:2015qia} in heavy ion collisions or using hydrodynamics in small systems \cite{Bozek:2013ska}. 

It was realized \cite{Ollitrault:1992bk} that there will be deviations of radial symmetry in the transverse flow due to anisotropic pressure gradients during the evolution. The Fourier decomposition of the angular distribution of particles with respect to the reaction plane has become the standard method with which to analyze flow anisotropies \cite{Voloshin:1994mz}. There are many experimental methods to measure those Fourier coefficients, like flow cumulants \cite{Bilandzic:2010jr}, but we shall not elaborate the discussion in this note.

Using Fourier decomposition, azimuthal dependence of the final hadron spectra becomes
\bge\label{eq:fourier}
E\frac{dN}{d^3p}=\frac{1}{\pt{}}\frac{dN}{d\pt{} dyd\phi}=\frac{1}{2\pi\pt{}}\frac{dN}{d\pt{} dy}\left(1+\sum_{n=1}^\infty 2v_{n}(\pt{})\cos(n\phi-\psi_n)\right),
\ende
where $v_n$ are the flow harmonics and $\psi_n$ event plane angles. The second harmonic coefficient, $v_2$, is called elliptic flow and it has a clear centrality dependence. Elliptic flow reflects the dominant almond-shaped geometry in the non-central collisions. The third harmonic, $v_3$, is called triangular flow. Triangular flow comes dominant in the ultra-central collisions and it is related to the fluctuations in the initial state of the collisions. For example, simultaneous description of the \pt{}\ and centrality dependence of the $v_2$ and $v_3$ coefficients puts constraints on the initial geometry fluctuations and the shear viscosity to entropy ratio $\eta/s$. A higher viscosity causes more rapid damping of the fluctuations ($v_3$) and slows the buildup of $v_2$. Based on these studies, it has been concluded that QGP shows features of nearly perfect liquid. \cite{Gale:2012rq, Song:2013qma}

The hydrodynamical models \cite{Gale:2012rq, Song:2013qma, Niemi:2015qia} go beyond the event averages of the flow coefficients (the $v_n$ in Eq.~(\ref{eq:fourier})) and study e.g. experimentally measured event-by-event fluctuations of flow \cite{Aad:2013xma} and correlations between event plane angles \cite{Aad:2014fla}. Such a global analysis provide significantly tighter constraints on QCD matter properties.

One may observe that there is no rapidity dependence in the flow coefficients in \eq{eq:fourier}. Consequently, flow leads into interesting long range correlations seen in the two-particle correlations. Given the single particle distribution (\ref{eq:fourier}), one obtains a similar Fourier decomposition for the two particle distribution \cite{Aamodt:2011by}
\bge\label{eq:2pleFourier}
\frac{dN^{\rm pair}}{d(\phi_t-\phi_a)}=:\frac{dN^{\rm pair}}{d\Delta\phi}\propto 1+\sum_{n=1}^\infty 2V_{n\Delta}(\pt{t},\pt{a})\cos(\Delta\phi),
\ende
where $\Delta\phi:=\phi_t-\phi_a$ is a difference in the azimuthal angle between trigger and associated hadron that have transverse momenta \pt{t}\ and \pt{a}, respectively. In the case of flow correlations only, one would have a connection to single particle flow coefficients
\bge\label{eq:flowFrag}
V_{n\Delta}(\pt{t},\pt{a})=v_n(\pt{t})v_n(\pt{a})
\ende
and you would see the correlations extending over large rapidity gaps $\Delta\eta:=\eta_t-\eta_a$. This gives an opportunity to determine flow coefficients for hadrons and also for identified flow \cite{ABELEV:2013wsa}: by choosing a symmetric bins for trigger and associated particles, $\pt{min}<\pt{a},\pt{t}<\pt{max}$, one finds
\bge\label{eq:2PC}
v^h_n\{2PC\}=\sqrt{V^{h-h}_{n\Delta}} \qquad v^i_n\{2PC\}=\frac{V^{h-i}_{n\Delta}}{\sqrt{V^{h-h}_{n\Delta}}},
\ende
where $i=\pi, \ {\rm K}, \ {\rm p}$ in case of identified flow. Above $h-h$ refers to unidentified charged hadron-hadron correlation in a symmetric \pt{}-bin and $h-i$ to a case where one of the hadrons is identified.

So far we have related the harmonics with collectivity in the system (hydrodynamical flow). However, one should be a bit careful since particularly hydrodynamics is expected to describe the bulk of the particle production, i.e. it would be valid description at $\pt{}<1\dots3\gevc$ depending on the particle species. However, as discussed in \sec{sec:RAA}, hard partons are not expected to thermalize but instead they traverse the medium losing their energy. But since the medium is expected to have a smaller geometrical extension in in-plane as compared to out-of-plane, the partons are (on the average) less suppressed at the in-plane direction, causing a positive second Fourier coefficient $v_2$. Hence, even though one generally speaks of ``flow coefficients'', the full range in \pt{}\ is not related with hydrodynamical flow.

The rest of this note concentrates on identified $v_2$ in \pbpb\ and \ppb\ collisions.

\subsection{Heavy ion collisions}\label{sec:v2-PbPb}

\begin{figure}[h]
\centering
\sidecaption
\includegraphics[width=7.0cm,clip]{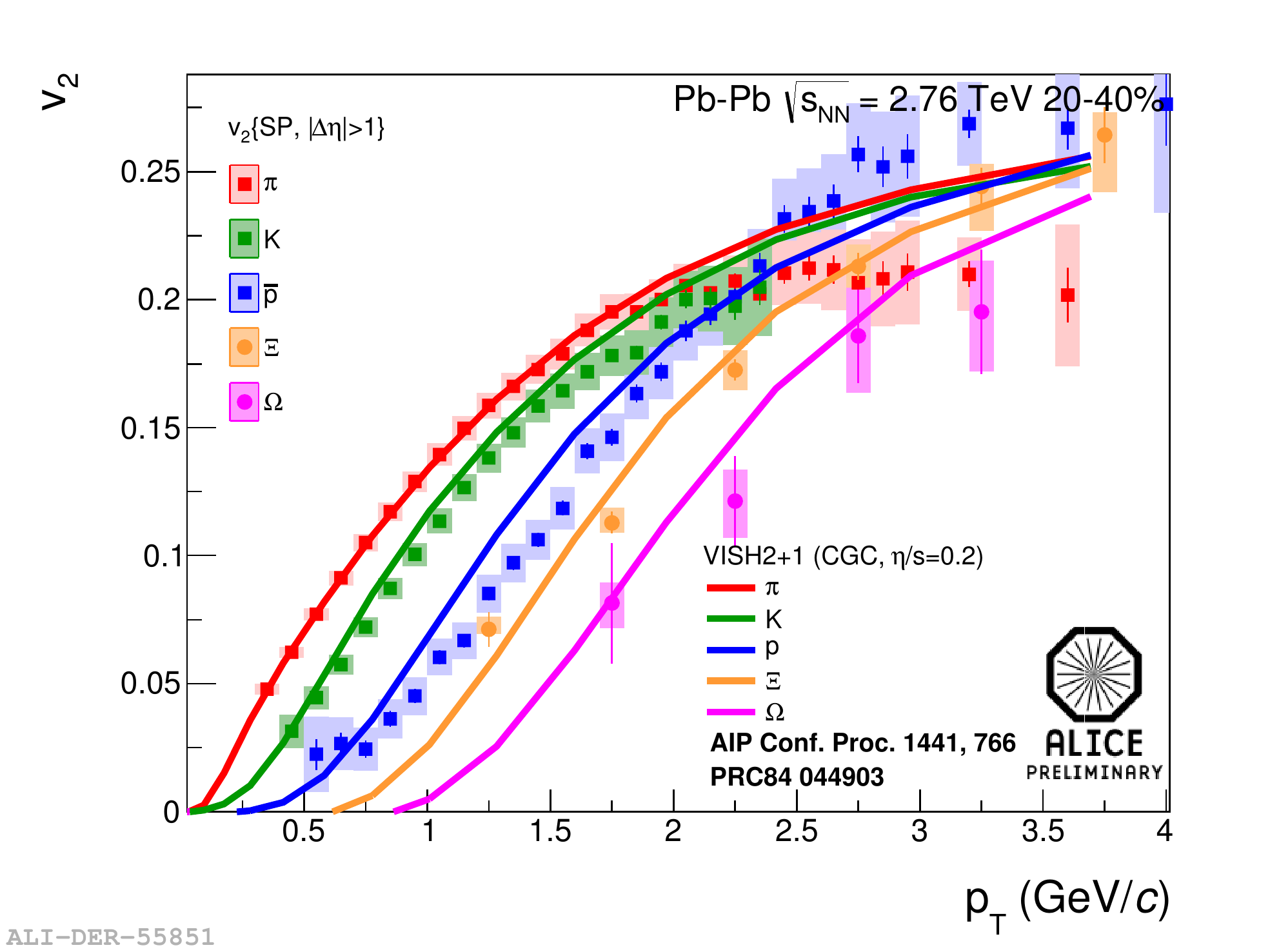}
\includegraphics[width=7.0cm,clip]{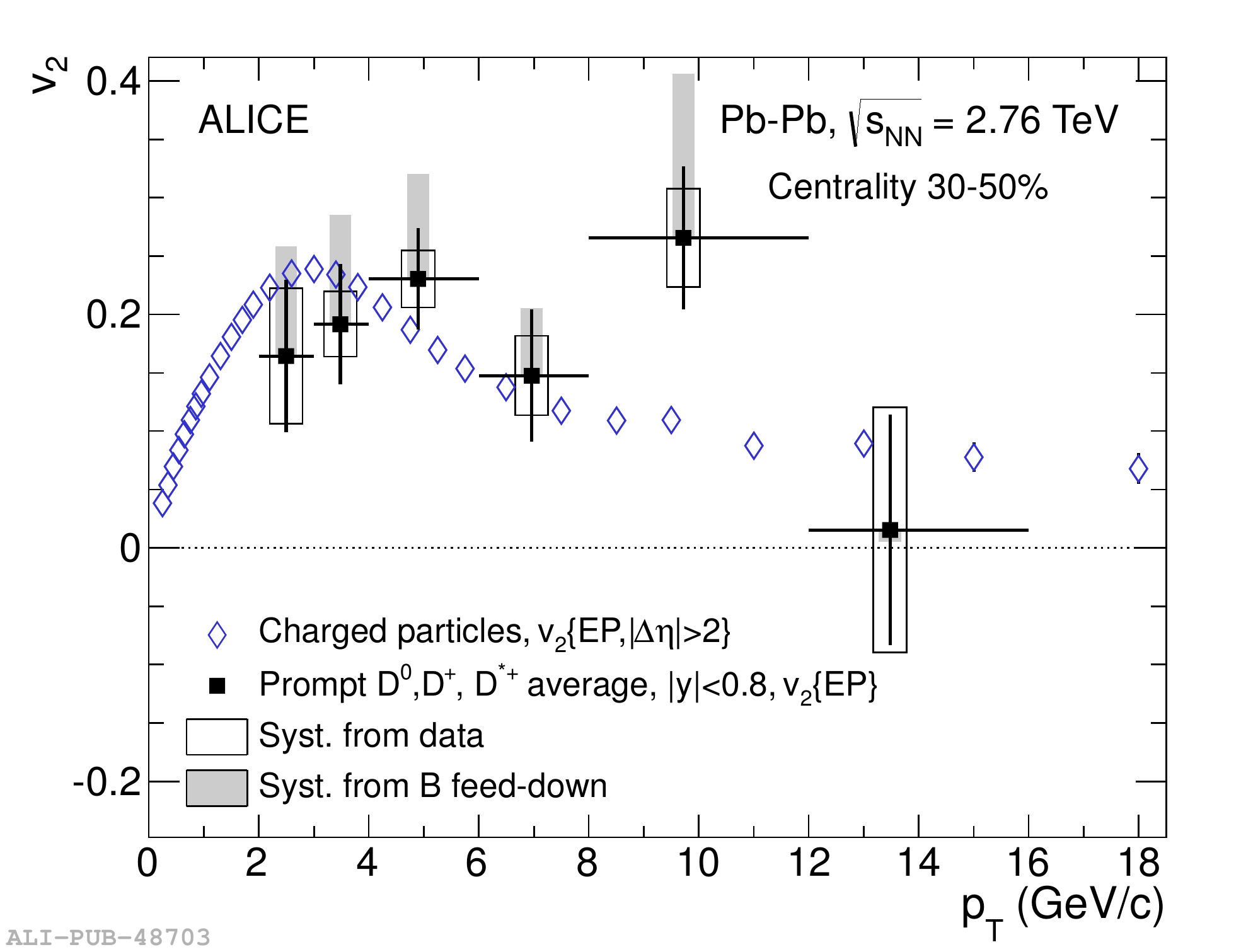}
\caption{Left: elliptic flow $v_2$ for identified particles in $\snn=2.76$\tev\ measured by ALICE \cite{Abelev:2014pua}. Right: D-meson and charged hadron $v_2$ compared \cite{Abelev:2013lca}.}
\label{fig:v2-identified}
\end{figure}
In the Blast Wave model (\ref{eq:BW}), radial flow breaks the $m_{\rm T}$--scaling present in the non-flowing case (i.e., if $\beta_{\rm max}=0$ then $({\rm yield})\sim K_1(-m_{\rm T}/T_{\rm kin})$). Heavier particles gain more momentum from the common collective flow field and this leads into mass ordering of elliptic flow at low-\pt{}. The left panel in \fig{fig:v2-identified} shows recent ALICE measurement\footnote{The published paper includes also the $v_2$ of ${\rm K}^0_{\rm s}$ and $\phi$ mesons, not shown in \fig{fig:v2-identified}.} \cite{Abelev:2014pua} for elliptic flow of $\pi^\pm, \ {\rm K}^\pm, \ {\rm p}+\bar{{\rm p}}, \Lambda+\bar\Lambda, \Xi+\bar\Xi$ and $\Omega+\bar\Omega$. One sees a clear mass ordering of the different hadrons and also that a sophisticated hydrodynamical simulation \cite{Shen:2011eg} can reproduce that qualitatively rather well, although a more detailed analysis presented in the ALICE paper shows some tensions. The study also provided evidence for constituent quark scaling \cite{Molnar:2003ff,Adare:2006ti,Abelev:2007qg} to be violated at the intermediate \pt{}\ region at the LHC as deviations up to $\sim20$~\% level are observed \cite{Abelev:2014pua}

The right panel in \fig{fig:v2-identified} shows the elliptic flow of average D-mesons compared to charged particle $v_2$. The measurements \cite{Abelev:2013lca} show that D-mesons have a positive $v_2$ in range $2<\pt{}<6\gevc$ interval with $5.7\sigma$ significance. This high-\pt{}\ range is outside the thermal region so these results, together with the D-meson \raa, provide very tight constraints on the heavy quark interactions with the medium.

\subsection{Double ridge in proton-lead collisions}\label{sec:flow-pPb}

An interesting observation in p-Pb collisions at the LHC was the appearance of a double ridge structure in two-particle correlations \cite{Abelev:2012ola}. Two-particle correlations are define as 
\bge
\frac{1}{N_{\rm trig}}\frac{d^2N}{d\Delta\phi d\Delta\eta},
\ende
where $\Delta\phi=\phi_t-\phi_a$ and $\Delta\eta=\eta_t-\eta_a$, such that trigger (t) and associated (a) transverse momenta are in the given bin, and also classify the events globally in terms of the charged particle multiplicity. Typically two-particle correlation functions have a near side peak at $(\Delta\phi,\Delta\eta)=(0,0)$ coming from hadrons associated with the fragmentation of the same jet as the trigger particle and an away-side peak at $\Delta\phi=\pi$, which is smeared in $\Delta\eta$ due to the different momentum fractions carried by partons in the initial hard scattering. If we now assume that the hard scattering is not significantly modified in proton-lead collision, we can assume that jet correlations from the hard scattering are similar in low- and high-multiplicity \ppb\ events. For the further support, we have checked that the correlation function in the low-multiplicity \ppb\ collisions is observed to be very similar with the one measured in proton-proton collisions. 
\begin{figure}[h]
\centering
\sidecaption
\includegraphics[width=5.0cm,clip]{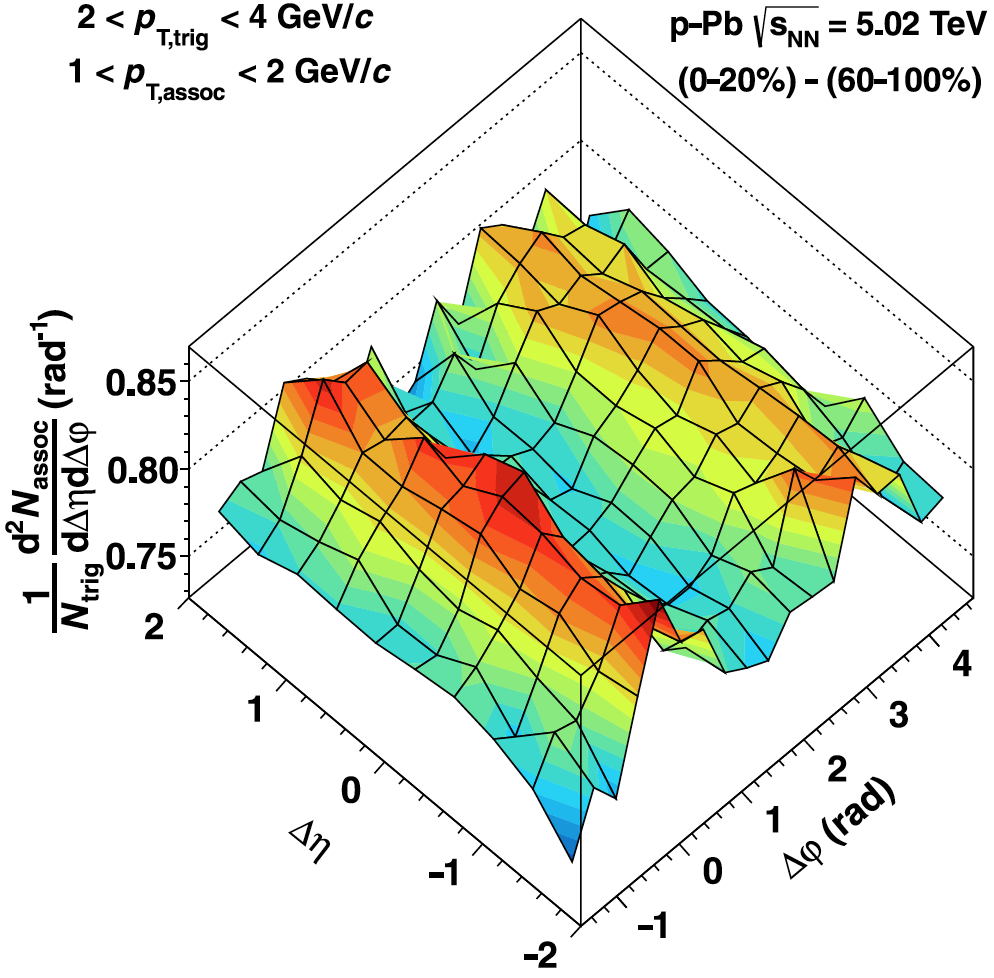}
\includegraphics[width=7.0cm,clip]{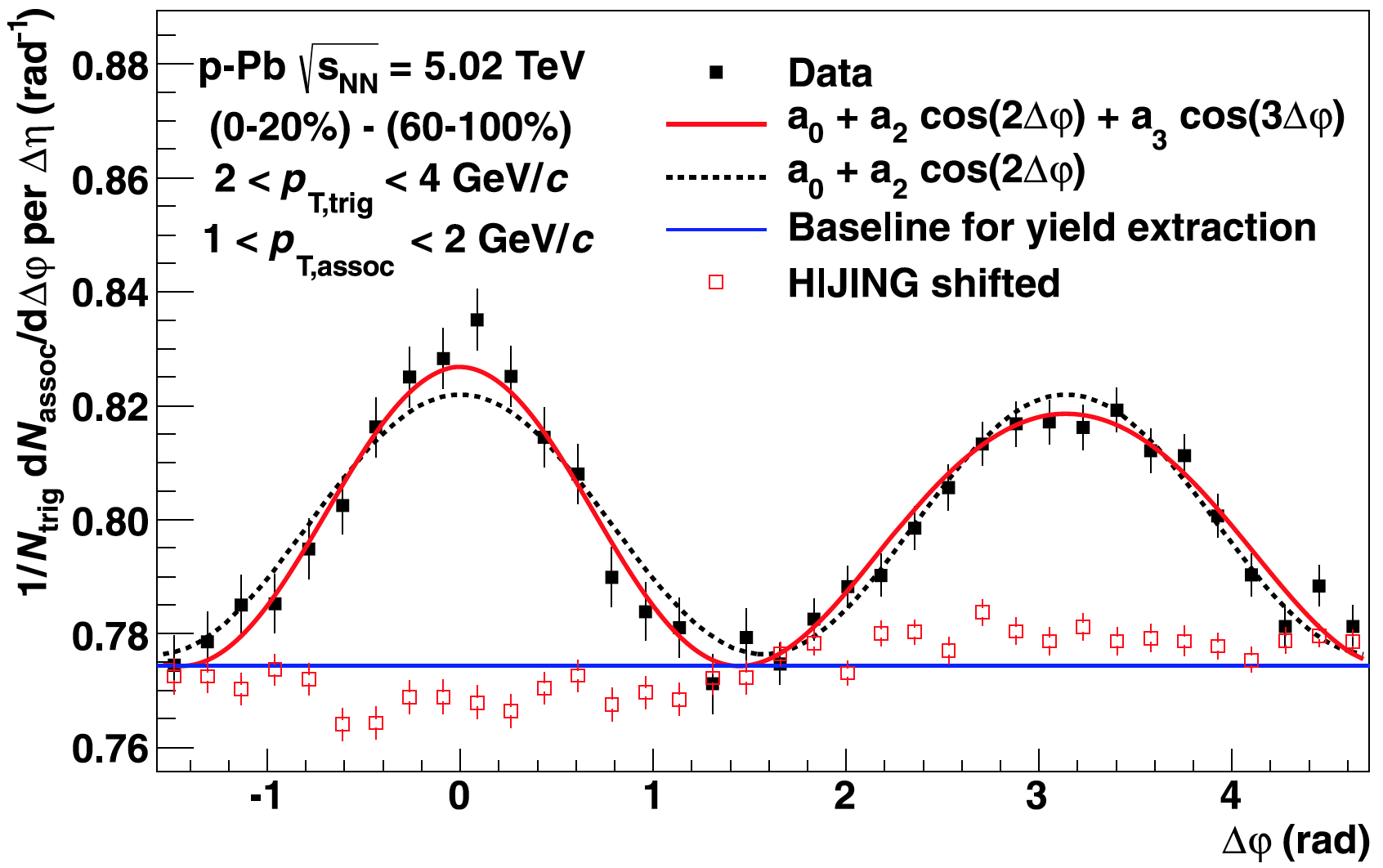}
\caption{Left: Associated yield per trigger particle in $\Delta\phi$ and $\Delta\eta$ for pairs of charged particles with $2<\pt{trig}<4$~\gevc\ and $1<\pt{assoc}<2$~\gevc\ in \ppb\ collisions at $\snn=5.02$~TeV for the 0–20\% multiplicity class,after subtraction of the associated yield obtained in the 60–100\% event class. Right: as left but projected onto $\Delta\phi$ averaged over $0.8<|\Delta\eta|<1.8$ on the near side and $|\Delta\eta|< 1.8$ on the away side. \cite{Abelev:2012ola}}
\label{fig:doubleRidge}
\end{figure}
Then we can study any residual correlations in high-multiplicity events, on top of the hard scattering, by subtracting the correlation functions measured at low-multiplicity from the correlation function measured in the high-multiplicity events. The left panel in \fig{fig:doubleRidge} shows the resulting residual correlation function, when you perform such a subtraction \cite{Abelev:2012ola}. One can observe a clear double ridge structure in $\Delta\phi$ that is elongated in $\Delta\eta$. Returning to the discussion of collective flow of single particles is reflected into two-particle correlations, see \eq{eq:2pleFourier}, we immediately realize that the observation very much resembles this expectation.

The right panel in \fig{fig:doubleRidge} shows the projection of the two-dimensional subtracted correlation function (left) into $\Delta\phi$ together with Fourier decomposition including $\cos(2\Delta\phi)$ and $\cos(3\Delta\phi)$ terms resembling elliptic and triangular flow in the single particle distributions. Note, however, that you cannot use directly \eq{eq:2PC} to obtain single particle $v_{2,3}$ here because the trigger and associated bins are asymmetric. Nevertheless, we observe that the double ridge structure is dominantly given by the 2$^{\rm nd}$ harmonic and after inclusion of 3$^{\rm rd}$ harmonic the description of the data already is very good. Results from HIJING simulation, that does not have any flow effects, is also shown and there is no similar double ridge structure seen indicating that this is not a trivial long range pseudorapidity correlation.

\begin{figure}[h]
\centering
\sidecaption
\includegraphics[width=8.0cm,clip]{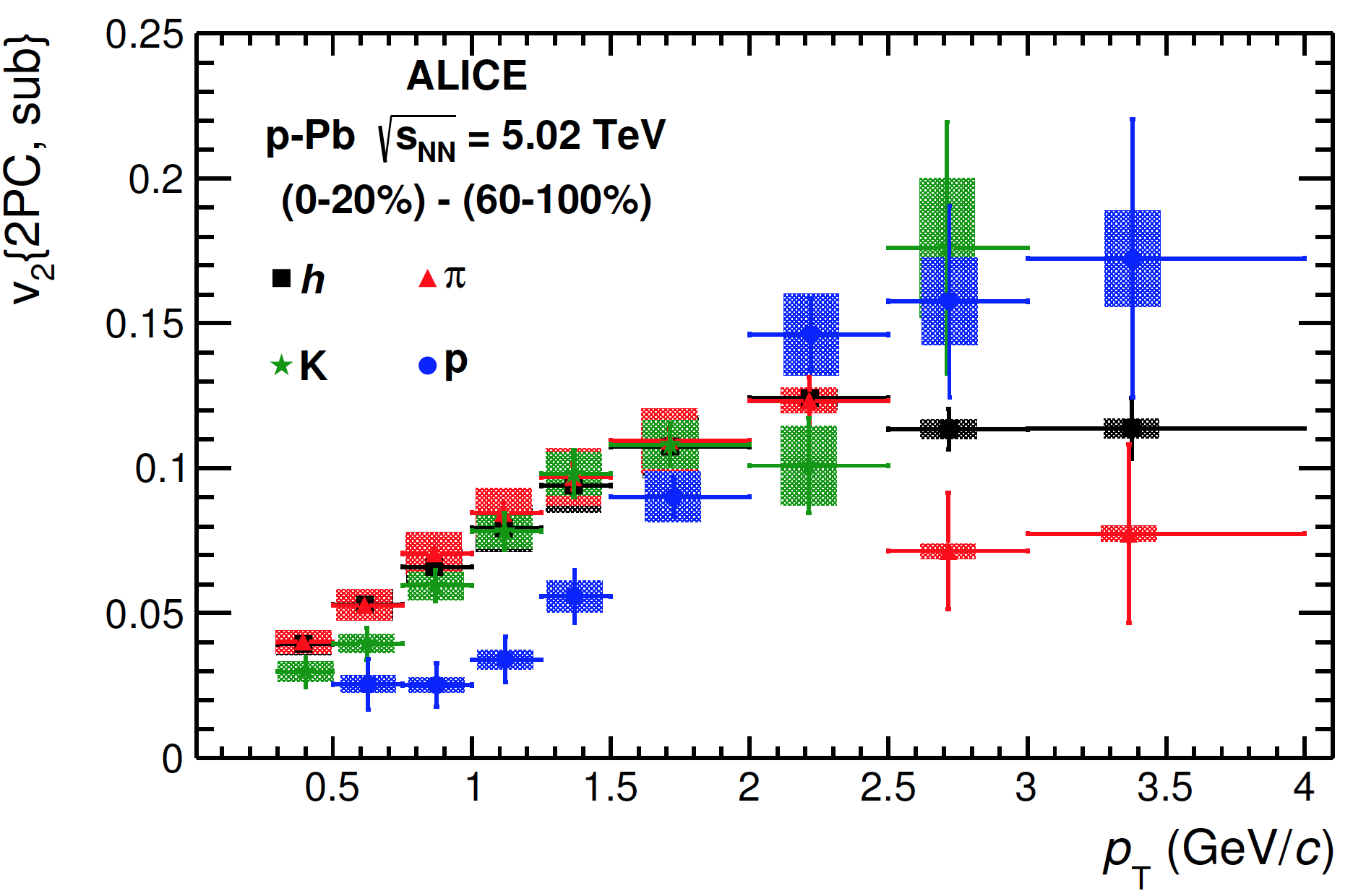}
\caption{Fourier coefficients $v_2\{ {\rm 2PC, sub} \}$ for pions, kaons and protons measured in the high-multiplicity \ppb\ collisions \cite{ABELEV:2013wsa}.}
\label{fig:v2-pPb}
\end{figure}
More detailed study of the properties of the observed double ridge was presented in \cite{ABELEV:2013wsa}. \fig{fig:v2-pPb} shows the $v_2\{{\rm 2PC, sub}\}$, extracted using \eq{eq:2PC}, where ``sub'' refers that the jet correlations are removed by subtracting the low-multiplicity correlation function from the high-multiplicity one before the projection to $\Delta\phi$. The observed $v_2$ values have a similar mass ordering than observed in \pbpb\ collisions, compare to left panel in \fig{fig:v2-identified}. While not yet conclusive proof, it shows that the Fourier analysis of the double ridge leads to a mass ordering in $v_2\{{\rm 2PC, sub}\}$ that is characteristic of collective flow. The ATLAS collaboration has also measured higher harmonics using two-particle correlations \cite{Aad:2014lta} and the CMS collaboration has also shown that multiparticle correlations \cite{Khachatryan:2015waa} give a non-zero $v_2$ at high-multiplicity proton-lead collisions. The latter strongly favour collective correlations.

Hydrodynamical calculations can qualitatively reproduce these observations \cite{Bozek:2013ska}. However, one might be concerned on validity of the (viscous) hydrodynamical description due to large pressure gradients in these small system \cite{Niemi:2014wta}.  An alternative Color Glass Condensate (CGC) explanation based on two gluon scattering from the same color field gives rise to similar correlations when the target gluon densities are high (saturation) \cite{Dusling:2013qoz}. It remains to be seen if the CGC based model can also explain the observed mass ordering.

\section{Summary}

ALICE is the dedicated heavy ion experiment at the CERN LHC that is equipped with excellent tracking and particle identification capabilities. Recent measurements have been presented of recent ALICE results for \raa\ and elliptic flow for identified particles in \pbpb\ and \ppb\ collisions. There are interesting hints for collective behaviour in high-multiplicity \ppb\ collisions. However, the effect of a Color Glass Condensate or other cold matter effects may yet provide an alternative explanation for the observed behaviour.

%
\bibliography{RASANEN_Sami_ICFNP2015}

\end{document}

%% file: defs.tex
\def\fig#1{Fig.~\ref{#1}}
\def\eq#1{Eq.~(\ref{#1})}
\def\sec#1{Sec.~\ref{#1}}

\def\bge{\begin{equation}}
\def\ende{\end{equation}}
\def\bgc{\begin{center}}
\def\endc{\end{center}}

\def\gev{\mbox{~GeV}}
\def\gevc{\mbox{~GeV/$c$}}

\def\tev{\mbox{~TeV}}

\def\la{\left<}
\def\ra{\right>}

\def\mean#1{\ensuremath{\la#1\ra}}

\def\sqrts{\ensuremath{\sqrt{s}}}

\def\pt#1{\ensuremath{p_{\rm T#1}}} 
\def\ptkv#1{\ensuremath{p^2_{\rm T#1}}}

\def\jpsi{\ensuremath{{\rm J}/\psi}}

\newcommand{\pp} {\ensuremath{\rm{pp}}}

\newcommand{\pbpb} {\ensuremath{\rm{Pb}-\rm{Pb}}}
\newcommand{\ppb} {\ensuremath{\rm{p}-\rm{Pb}}}

\newcommand{\raa} {\ensuremath{R_{AA}}}
\newcommand{\rpa} {\ensuremath{R_{{\rm p}A}}}

\newcommand{\snn} {\ensuremath{\sqrt{s_{\rm NN}}}}

\def\lsim{\raise0.3ex\hbox{$<$\kern-0.75em\raise-1.1ex\hbox{$\sim$}}}
\def\gsim{\raise0.3ex\hbox{$>$\kern-0.75em\raise-1.1ex\hbox{$\sim$}}}

%% file: RASANEN_Sami_ICFNP2015.bbl
\begin{thebibliography}{70}

\bibitem{Akiba:2015jwa}
Y.~Akiba et~al. (2015), \texttt{1502.02730}

\bibitem{smallSystems}
I.~Tserruya, These proceedings  (2015)

\bibitem{Adare:2014keg}
A.~Adare et~al. (PHENIX), Phys. Rev. Lett. \textbf{114}, 192301 (2015),
  \texttt{1404.7461}

\bibitem{Adamczyk:2015xjc}
L.~Adamczyk et~al. (STAR), Phys. Lett. \textbf{B747}, 265 (2015),
  \texttt{1502.07652}

\bibitem{Adare:2015ctn}
A.~Adare et~al. (PHENIX), Phys. Rev. Lett. \textbf{115}, 142301 (2015),
  \texttt{1507.06273}

\bibitem{Abelev:2012ola}
B.~Abelev et~al. (ALICE), Phys. Lett. \textbf{B719}, 29 (2013),
  \texttt{1212.2001}

\bibitem{Aad:2012gla}
G.~Aad et~al. (ATLAS), Phys. Rev. Lett. \textbf{110}, 182302 (2013),
  \texttt{1212.5198}

\bibitem{Chatrchyan:2013nka}
S.~Chatrchyan et~al. (CMS), Phys. Lett. \textbf{B724}, 213 (2013),
  \texttt{1305.0609}

\bibitem{Khachatryan:2010gv}
V.~Khachatryan et~al. (CMS), JHEP \textbf{09}, 091 (2010), \texttt{1009.4122}

\bibitem{Aad:2015gqa}
G.~Aad et~al. (ATLAS) (2015), \texttt{1509.04776}

\bibitem{Khachatryan:2015lva}
V.~Khachatryan et~al. (CMS) (2015), \texttt{1510.03068}

\bibitem{Aamodt:2008zz}
K.~Aamodt et~al. (ALICE), JINST \textbf{3}, S08002 (2008)

\bibitem{Aad:2008zzm}
G.~Aad et~al. (ATLAS), JINST \textbf{3}, S08003 (2008)

\bibitem{Chatrchyan:2008aa}
S.~Chatrchyan et~al. (CMS), JINST \textbf{3}, S08004 (2008)

\bibitem{Adam:2015pna}
J.~Adam et~al. (ALICE), Nature Phys.  (2015), \texttt{1508.03986}

\bibitem{Adam:2015qaa}
J.~Adam et~al. (ALICE), Eur. Phys. J. \textbf{C75}, 226 (2015),
  \texttt{1504.00024}

\bibitem{Abelev:2014ypa}
B.B. Abelev et~al. (ALICE), Eur. Phys. J. \textbf{C74}, 3108 (2014),
  \texttt{1405.3794}

\bibitem{ALICE:2014dla}
B.B. Abelev et~al. (ALICE), Phys. Rev. \textbf{D91}, 112012 (2015),
  \texttt{1411.4969}

\bibitem{Wang:1998hs}
X.N. Wang, Phys. Rev. Lett. \textbf{81}, 2655 (1998), \texttt{hep-ph/9804384}

\bibitem{Adcox:2001jp}
K.~Adcox et~al. (PHENIX), Phys. Rev. Lett. \textbf{88}, 022301 (2002),
  \texttt{nucl-ex/0109003}

\bibitem{Aamodt:2010cz}
K.~Aamodt et~al. (ALICE), Phys. Rev. Lett. \textbf{106}, 032301 (2011),
  \texttt{1012.1657}

\bibitem{Antreasyan:1978cw}
D.~Antreasyan, J.W. Cronin, H.J. Frisch, M.J. Shochet, L.~Kluberg, P.A. Piroue,
  R.L. Sumner, Phys. Rev. \textbf{D19}, 764 (1979)

\bibitem{Eskola:2009uj}
K.J. Eskola, H.~Paukkunen, C.A. Salgado, JHEP \textbf{04}, 065 (2009),
  \texttt{0902.4154}

\bibitem{Adam:2015ewa}
J.~Adam et~al. (ALICE), Phys. Lett. \textbf{B746}, 1 (2015),
  \texttt{1502.01689}

\bibitem{jetTalk}
A.~Shabetai, These proceedings  (2015)

\bibitem{Aamodt:2010jd}
K.~Aamodt et~al. (ALICE), Phys. Lett. \textbf{B696}, 30 (2013),
  \texttt{1012.1004}

\bibitem{Aad:2015wga}
G.~Aad et~al. (ATLAS), JHEP \textbf{09}, 050 (2015), \texttt{1504.04337}

\bibitem{CMS:2012aa}
S.~Chatrchyan et~al. (CMS), Eur. Phys. J. \textbf{C72}, 1945 (2012),
  \texttt{1202.2554}

\bibitem{Adam:2015kca}
J.~Adam et~al. (ALICE) (2015), \texttt{1506.07287}

\bibitem{Fries:2003kq}
R.J. Fries, B.~Muller, C.~Nonaka, S.A. Bass, Phys. Rev. \textbf{C68}, 044902
  (2003), \texttt{nucl-th/0306027}

\bibitem{HFtheory}
M.~Nardi, These proceedings  (2015)

\bibitem{HFexp}
M.J. Kweon, These proceedings  (2015)

\bibitem{Dokshitzer:2001zm}
Y.L. Dokshitzer, D.E. Kharzeev, Phys. Lett. \textbf{B519}, 199 (2001),
  \texttt{hep-ph/0106202}

\bibitem{Matsui:1986dk}
T.~Matsui, H.~Satz, Phys. Lett. \textbf{B178}, 416 (1986)

\bibitem{Arnaldi:2009ph}
R.~Arnaldi (NA60), Nucl. Phys. \textbf{A830}, 345C (2009), \texttt{0907.5004}

\bibitem{Adare:2007gn}
A.~Adare et~al. (PHENIX), Phys. Rev. \textbf{C77}, 024912 (2008), [Erratum:
  Phys. Rev.C79,059901(2009)], \texttt{0903.4845}

\bibitem{Chatrchyan:2012lxa}
S.~Chatrchyan et~al. (CMS), Phys. Rev. Lett. \textbf{109}, 222301 (2012),
  \texttt{1208.2826}

\bibitem{Abelev:2012rv}
B.~Abelev et~al. (ALICE), Phys. Rev. Lett. \textbf{109}, 072301 (2012),
  \texttt{1202.1383}

\bibitem{Abelev:2013ila}
B.B. Abelev et~al. (ALICE), Phys. Lett. \textbf{B734}, 314 (2014),
  \texttt{1311.0214}

\bibitem{Adam:2015rba}
J.~Adam et~al. (ALICE), JHEP \textbf{07}, 051 (2015), \texttt{1504.07151}

\bibitem{Zhou:2014kka}
K.~Zhou, N.~Xu, Z.~Xu, P.~Zhuang, Phys. Rev. \textbf{C89}, 054911 (2014),
  \texttt{1401.5845}

\bibitem{Zhao:2011cv}
X.~Zhao, R.~Rapp, Nucl. Phys. \textbf{A859}, 114 (2011), \texttt{1102.2194}

\bibitem{ALICE:2012mj}
B.~Abelev et~al. (ALICE), Phys. Rev. Lett. \textbf{110}, 082302 (2013),
  \texttt{1210.4520}

\bibitem{Abelev:2014hha}
B.B. Abelev et~al. (ALICE), Phys. Rev. Lett. \textbf{113}, 232301 (2014),
  \texttt{1405.3452}

\bibitem{Helenius:2012wd}
I.~Helenius, K.J. Eskola, H.~Honkanen, C.A. Salgado, JHEP \textbf{07}, 073
  (2012), \texttt{1205.5359}

\bibitem{Tribedy:2011aa}
P.~Tribedy, R.~Venugopalan, Phys. Lett. \textbf{B710}, 125 (2012), [Erratum:
  Phys. Lett.B718,1154(2013)], \texttt{1112.2445}

\bibitem{Schnedermann:1993ws}
E.~Schnedermann, J.~Sollfrank, U.W. Heinz, Phys. Rev. \textbf{C48}, 2462
  (1993), \texttt{nucl-th/9307020}

\bibitem{Abelev:2013vea}
B.~Abelev et~al. (ALICE), Phys. Rev. \textbf{C88}, 044910 (2013),
  \texttt{1303.0737}

\bibitem{Eskola:2002wx}
K.J. Eskola, H.~Niemi, P.V. Ruuskanen, S.S. Rasanen, Phys. Lett. \textbf{B566},
  187 (2003), \texttt{hep-ph/0206230}

\bibitem{Song:2013qma}
H.~Song, S.~Bass, U.W. Heinz, Phys. Rev. \textbf{C89}, 034919 (2014),
  \texttt{1311.0157}

\bibitem{Gale:2012rq}
C.~Gale, S.~Jeon, B.~Schenke, P.~Tribedy, R.~Venugopalan, Phys. Rev. Lett.
  \textbf{110}, 012302 (2013), \texttt{1209.6330}

\bibitem{Niemi:2015qia}
H.~Niemi, K.J. Eskola, R.~Paatelainen (2015), \texttt{1505.02677}

\bibitem{Bozek:2013ska}
P.~Bozek, W.~Broniowski, G.~Torrieri, Phys. Rev. Lett. \textbf{111}, 172303
  (2013), \texttt{1307.5060}

\bibitem{Ollitrault:1992bk}
J.Y. Ollitrault, Phys. Rev. \textbf{D46}, 229 (1992)

\bibitem{Voloshin:1994mz}
S.~Voloshin, Y.~Zhang, Z. Phys. \textbf{C70}, 665 (1996),
  \texttt{hep-ph/9407282}

\bibitem{Bilandzic:2010jr}
A.~Bilandzic, R.~Snellings, S.~Voloshin, Phys. Rev. \textbf{C83}, 044913
  (2011), \texttt{1010.0233}

\bibitem{Aad:2013xma}
G.~Aad et~al. (ATLAS), JHEP \textbf{11}, 183 (2013), \texttt{1305.2942}

\bibitem{Aad:2014fla}
G.~Aad et~al. (ATLAS), Phys. Rev. \textbf{C90}, 024905 (2014),
  \texttt{1403.0489}

\bibitem{Aamodt:2011by}
K.~Aamodt et~al. (ALICE), Phys. Lett. \textbf{B708}, 249 (2012),
  \texttt{1109.2501}

\bibitem{ABELEV:2013wsa}
B.B. Abelev et~al. (ALICE), Phys. Lett. \textbf{B726}, 164 (2013),
  \texttt{1307.3237}

\bibitem{Abelev:2014pua}
B.B. Abelev et~al. (ALICE), JHEP \textbf{06}, 190 (2015), \texttt{1405.4632}

\bibitem{Abelev:2013lca}
B.~Abelev et~al. (ALICE), Phys. Rev. Lett. \textbf{111}, 102301 (2013),
  \texttt{1305.2707}

\bibitem{Shen:2011eg}
C.~Shen, U.~Heinz, P.~Huovinen, H.~Song, Phys. Rev. \textbf{C84}, 044903
  (2011), \texttt{1105.3226}

\bibitem{Molnar:2003ff}
D.~Molnar, S.A. Voloshin, Phys. Rev. Lett. \textbf{91}, 092301 (2003),
  \texttt{nucl-th/0302014}

\bibitem{Adare:2006ti}
A.~Adare et~al. (PHENIX), Phys. Rev. Lett. \textbf{98}, 162301 (2007),
  \texttt{nucl-ex/0608033}

\bibitem{Abelev:2007qg}
B.I. Abelev et~al. (STAR), Phys. Rev. \textbf{C75}, 054906 (2007),
  \texttt{nucl-ex/0701010}

\bibitem{Aad:2014lta}
G.~Aad et~al. (ATLAS), Phys. Rev. \textbf{C90}, 044906 (2014),
  \texttt{1409.1792}

\bibitem{Khachatryan:2015waa}
V.~Khachatryan et~al. (CMS), Phys. Rev. Lett. \textbf{115}, 012301 (2015),
  \texttt{1502.05382}

\bibitem{Niemi:2014wta}
H.~Niemi, G.S. Denicol (2014), \texttt{1404.7327}

\bibitem{Dusling:2013qoz}
K.~Dusling, R.~Venugopalan, Phys. Rev. \textbf{D87}, 094034 (2013),
  \texttt{1302.7018}

\end{thebibliography}
